\documentclass[fleqn,usenatbib]{mnras}

\usepackage{newtxtext,newtxmath}

\usepackage[T1]{fontenc}

\DeclareRobustCommand{\VAN}[3]{#2}
\let\VANthebibliography\thebibliography
\def\thebibliography{\DeclareRobustCommand{\VAN}[3]{##3}\VANthebibliography}

\usepackage{tikz}
\usepackage{tikzscale}
\usepackage{graphicx}	
\usepackage{amsmath}	
\usepackage{bm}
\usepackage{booktabs}
\usepackage{fontawesome5}
\usepackage{hyperref}
\usepackage{array}
\usepackage{caption}
\usepackage{graphicx}
\usepackage{siunitx}
\usepackage[normalem]{ulem}
\usepackage{colortbl}
\usepackage{multirow}
\usepackage{hhline}
\usepackage{calc}
\usepackage{tabularx}
\usepackage{threeparttable}
\usepackage{wrapfig}
\usepackage{adjustbox}
\usepackage[english]{babel}
\usepackage[autostyle, english = american]{csquotes}
\MakeOuterQuote{"}
\hypersetup{
    colorlinks=true, 
    urlcolor=cyan,
    }

\newcommand{\github}[1]{%
   \href{#1}{\faGithubSquare}%
}
\newcommand{\lmstar}{$\log_{10}(\mathrm{M}_*/\mathrm{M}_{\odot})$}
\newcommand{\logenv}{$\log_{10}(\Sigma_5/\mathrm{Mpc}^{-2})$}
\newcommand{\logsfr}{$\log_{10}(\mathrm{SFR}/\mathrm{M}_{\odot} \mathrm{yr}^{-1})$}
\newcommand{\logre}{$\log_{10}(r_e/\mathrm{kpc})$}
\newcommand{\Msun}{M$_{\odot}$}

\title[Predicting Kinematic Morphology]{The SAMI galaxy survey: predicting kinematic morphology with logistic regression}

\author[Sam P. Vaughan et al.]{Sam P. Vaughan,$^{1, 2, 3, 4}$\thanks{E-mail: sam.vaughan@mq.edu.au (SPV)}
Jesse van de Sande$^{5, 4}$,
A. Fraser-McKelvie$^{6, 4}$,
Scott Croom$^{5, 4}$,
Richard McDermid$^{1,4}$,
\newauthor Benoit Liquet-Weiland$^{1, 7}$, Stefania Barsanti$^{8, 4}$, Luca Cortese$^{6,4}$, Sarah Brough$^{9,4}$, Sarah Sweet$^{10,4}$, 
\newauthor Julia J. Bryant$^{5, 4}$, Michael Goodwin$^{11}$ and Jon Lawrence$^{12}$\\
$^{1}$School of Mathematical and Physical Sciences, Macquarie University, NSW 2109, Australia\\ 
$^{2}$Astronomy, Astrophysics and Astrophotonics Research Centre, Macquarie University, Sydney, NSW 2109, Australia\\
$^{3}$Centre for Astrophysics and Supercomputing, School of Science, Swinburne University of Technology, Hawthorn, VIC 3122, Australia\\
$^{4}$ARC Centre of Excellence for All Sky Astrophysics in 3 Dimensions (ASTRO 3D), Australia\\
$^{5}$Sydney Institute for Astronomy, School of Physics, A28, The University of Sydney, NSW, 2006\\
$^{6}$International Centre for Radio Astronomy Research, The University of Western Australia, 35 Stirling Highway, Crawley WA 6009, Australia\\
$^{7}$Laboratoire de Mathématiques et de leurs Applications, Université de Pau et des Pays de l'Adour, E2S UPPA, CNRS, Anglet, France.\\
$^{8}$Research School of Astronomy and Astrophysics, The Australian National University, Canberra, ACT 2611, Australia\\
$^{9}$School of Physics, University of New South Wales, NSW 2052, Australia\\
$^{10}$School of Mathematics and Physics, University of Queensland, Brisbane, QLD 4072, Australia\\
$^{11}$AAO-MQ, Faculty of Science and Engineering, Macquarie University, 105 Delhi Rd, North Ryde, NSW 2113, Australia\\
$^{12}$Australian Astronomical Optics - Macquarie, Macquarie University, NSW 2109, Australia\\
}

\date{Accepted 2024 February 5. Received 2024 February 1; in original form 2023 October 18}

\pubyear{2023}

\begin{document}
\pagerange{?--?}
\maketitle

\begin{abstract}
    We use the SAMI galaxy survey to study the the kinematic morphology-density relation: the observation that the fraction of slow rotator galaxies increases towards dense environments. We build a logistic regression model to quantitatively study the dependence of kinematic morphology (whether a galaxy is a fast rotator or slow rotator) on a wide range of parameters, without resorting to binning the data. Our model uses a combination of stellar mass, star-formation rate (SFR), $r$-band half-light radius and a binary variable based on whether the galaxy's observed ellipticity ($\epsilon$) is less than 0.4. We show that, at fixed mass, size, SFR and $\epsilon$, a galaxy's local environmental surface density (\logenv{}) gives no further information about whether a galaxy is a slow rotator, i.e. the observed kinematic-morphology density relation can be entirely explained by the well-known correlations between environment and other quantities. We show how our model can be applied to different galaxy surveys to predict the fraction of slow rotators which would be observed and discuss its implications for the formation pathways of slow rotators.
\end{abstract}




\section{Introduction}

Large galaxy surveys of the local Universe using Integral Field Spectrographs are now commonplace. Arguably, the most important results to have come from such surveys has been in their application to galaxy dynamics. Using data from projects such as SAMI \citep{Croom:2012,Bryant:2015}, CALIFA \citep{Sanchez:2012} and MaNGA \citep{Bundy:2015}, and building on previous work such as ATLAS3D \citep{Cappellari:2011}, SAURON \citep{Bacon:2001} and others \citep[e.g.][]{Davies:1983}, astronomers have revealed a dichotomy between two classes of galaxies known as "Fast Rotators" and "Slow Rotators" (e.g. \citealt{Cappellari:2016}).

The majority of galaxies in the local volume are found to be Fast Rotators (FRs), which make up $\approx80$\% of the stellar mass budget of nearby galaxies \citep{Guo:2020,Fraser-McKelvie:2022}. Their dynamics are dominated by ordered rotation, implying the presence of rotating stellar discs and axisymmetric intrinsic shapes. As a function of increasing bulge fraction, the locus of FRs on the mass-size plane joins neatly to the population of spiral galaxies \citep{Cappellari:2011, Kormendy:2012, Cappellari:2013}, hinting at a common evolutionary pathway.

Slow Rotators (SRs), on the other hand, are fundamentally different objects (e.g. \citealt{Penoyre:2017}). Their kinematics are clearly inconsistent with simple stellar discs, instead showing evidence of kinematically decoupled cores, complex and irregular orbits or no net rotation at all. Their intrinsic shapes are (weakly) triaxial, and their observed apparent ellipticities are always rounder than $\epsilon\approx0.4$ \citep{Cappellari:2016}.

Since the discovery of this dichotomy, a number of studies have investigated the role of galaxy environment on the presence of SRs. From a theoretical and observational perspective, a trend with environment might be expected. According to the $\Lambda$CDM cosmological paradigm, galaxies reside in dark-matter halos which grow from overdensities in the primordial matter distribution. It has been shown in simulations that the properties of these halos depends on environment, such that a dark matter halo in a void evolves with different properties to a sub-halo of a galaxy cluster \citep{Avila-Reese:2005}. Halos in dense regions of the Universe also grow more rapidly and form earlier than halos in regions of average density \citep{Gao:2005, Maulbetsch:2007}.

Furthermore, there are a number of well-known observed correlations between a galaxy's properties and its local environment. The (visual) morphology-density relation \citep{Dressler:1980,Dressler:1997} shows that the fraction of early type galaxies increases and the fraction of spiral galaxies decrease in a population as the local environmental density increases. Galaxies in dense environments also tend to have redder colours \citep[e.g.][]{Pimbblet:2002, Balogh:2004, Bamford:2009, Pandey:2020}, are less likely to contain emission lines in their spectra \citep{Gisler:1978} and show suppressed star-formation rates \citep{Lewis:2002, Gomez:2003, Calvi:2018}.

A number of previous studies have measured the fraction of galaxies classified as SRs as a function of local density, finding that the fraction of SRs increases towards denser environments \citep{Cappellari:2011b, Houghton:2013, D'Eugenio:2013, Scott:2014, Fogarty:2014}. However, it has also been suggested that the underlying correlation between mass and environment is able to explain this observation, such that the kinematic morphology-density relation does not exist at fixed stellar mass (\citealt{Veale:2017, Brough:2017, Greene:2017}). This conclusion is not universally accepted, however: conversely, \citet{Scott:2014}, \citet{Wang:2020} and \citet{vandeSande:2021} do find a weak dependence on environment at fixed stellar mass.

Previous literature on the topic approaches the problem in the same way: by stratifying observations into bins of varying mass and environmental density and measuring the fraction of SRs in each bin. The drawback to this technique is the difficulty in expanding the analysis to include further variables of interest, as attempting to bin observations in more than two dimensions dramatically increases the survey size required.

This work takes a different approach. By using logistic regression, a generalised linear model appropriate when observations are binary (i.e. whether a galaxy is a FR or SR), we investigate the dependence of kinematic morphology on a number of different galaxy properties without resorting to binning. In particular, we attempt to quantitatively study the role of environment on kinematic morphology.

This paper is organised as follows. Section \ref{sec:sample_selection} outlines the sample of SAMI galaxies used in this work. Section \ref{sec:logistic_regression} gives a short introduction to logistic regression and outlines our modelling procedure. Section \ref{sec:results} describes our results. We discuss these results in Section \ref{sec:discussion} and present our conclusions in Section \ref{sec:conclusion}. 
\section{Sample Selection}
\label{sec:sample_selection}

The data used in the study come from the SAMI Galaxy Survey, described below in Section \ref{sec:SAMI_galaxy_survey}. In particular, we use measurements of kinematic morphology for each galaxy (described in Section \ref{sec:kinematic_measurements}), as well as ancillary measurements such as star-formation rates, stellar masses and a measure of environmental density (described in Section \ref{sec:ancillary}). The SAMI Galaxy Survey is the only large galaxy survey which combines the spatially-resolved spectroscopy necessary to measure kinematic morphology with a dedicated cluster survey to sample a wide range of environmental densities.

\begin{figure*}
    \includegraphics[width=0.45\textwidth]{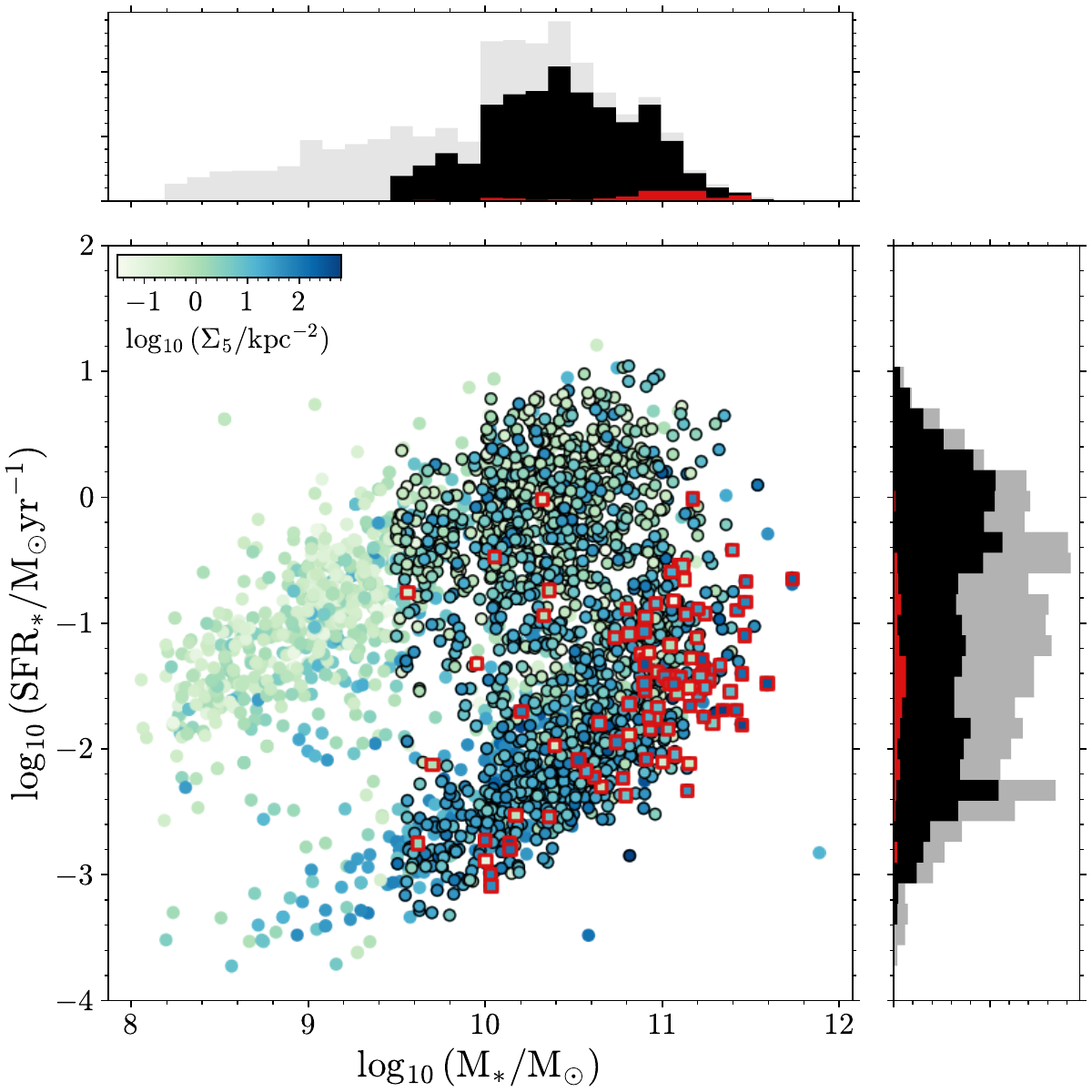}~
    \includegraphics[width=0.45\textwidth]{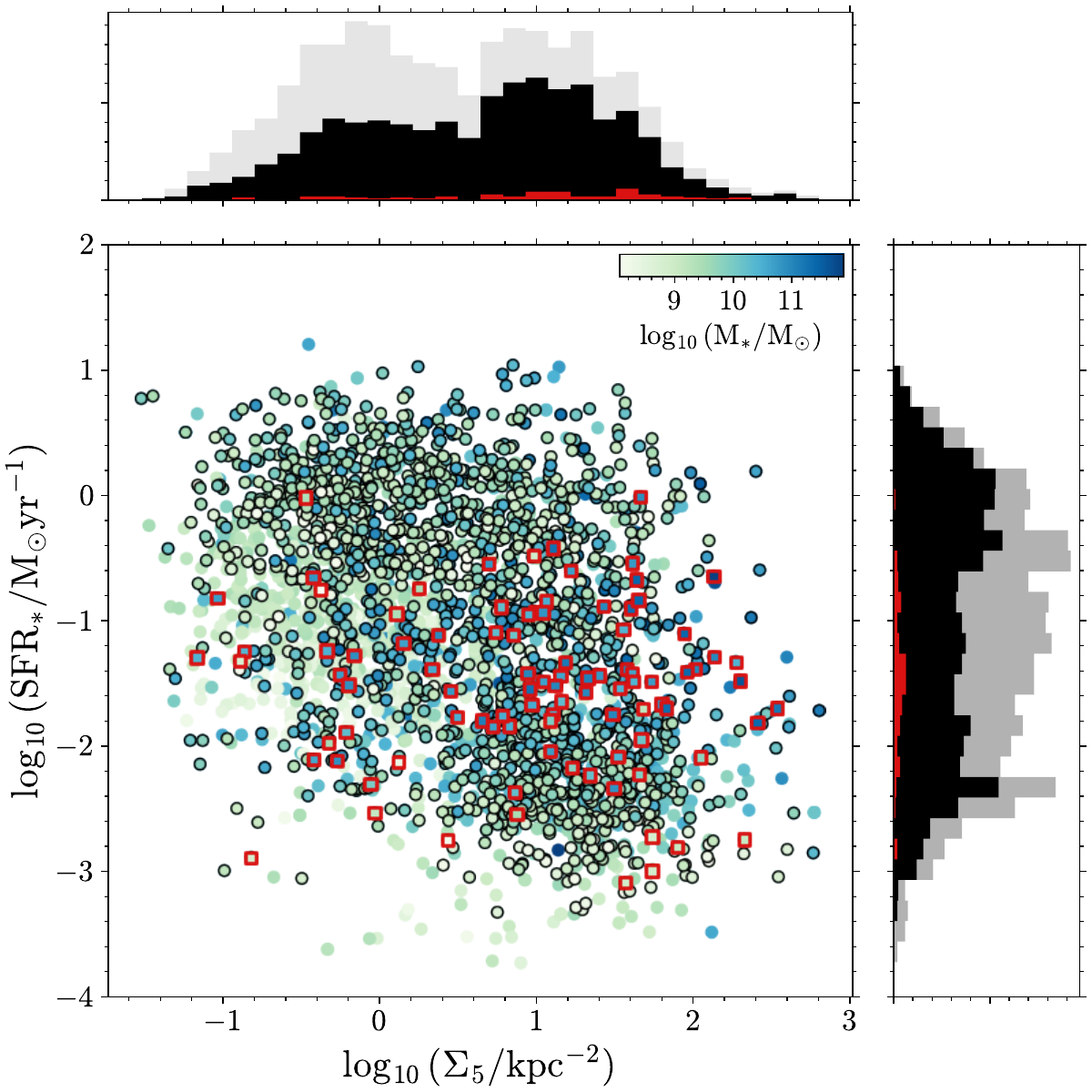}\\
    \includegraphics[width=0.99\textwidth]{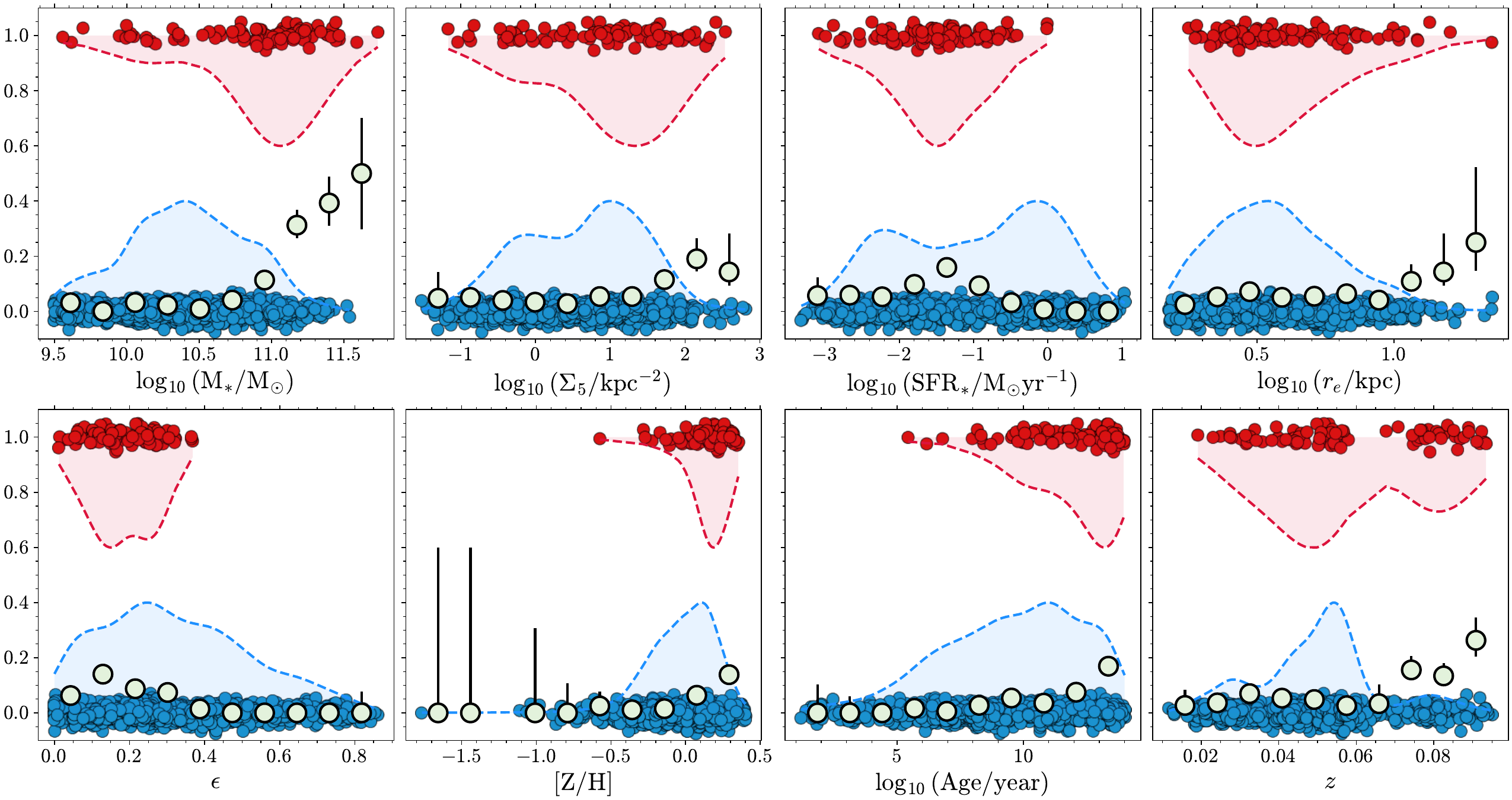}
    \caption{The sample of galaxies used in this paper. \textbf{Top}: galaxies in the full sample are shown as coloured points. Galaxies which are classified as FRs are shown outlined in black. Galaxies classified as SRs are shown as squares outlined in red. Those without an outline do not have a reliable kinematic classification. Histograms show the marginal distributions of each property, with the full sample coloured grey, fast rotators in black and slow-rotators in red. The left-hand panel shows stellar mass plotted against star-formation rate, with points coloured by their local environmental surface density, $\log_{10}(\Sigma_5)$. The right hand panel shows star-formation rate against $\log_{10}(\Sigma_5)$, with points coloured by stellar mass. \textbf{Bottom}: the kinematic classification of galaxies in the sample as a function of stellar mass, environmental surface density, star-formation rate and $r$-band half-light radius (top row); and ellipticity, mass-weighted metallicity within $r_e$, mass-weighted age within $r_e$ and spectroscopic redshift (bottom row). Slow rotators (SRs) are plotted at $y=1$ and fast rotators (FRs) are plotted at $y=0$. A small random jitter has been added to each point to avoid them being plotted on top of each other. The white points with error bars show the fraction of SRs in bins of the $x$-axis variable, with the error-bars representing the 16\textsuperscript{th} and 84\textsuperscript{th} percentiles of a Binomial distribution. Red and blue shaded regions show one-dimensional kernel density estimates of the distribution of slow and fast rotators as a function of each $x$ axis variable.}
    \label{fig:sample}
\end{figure*}

\subsection{The SAMI Galaxy Survey}
\label{sec:SAMI_galaxy_survey}

We use observations from the Data Release 3 of the SAMI Galaxy Survey \citep{Croom:2021}. The Sydney-AAO Multi-object Integral field spectrograph (SAMI; \citealt{Croom:2012}) is mounted at the prime focus on the Anglo-Australian Telescope on a top-end which provides a 1 degree diameter field of view. SAMI uses 13 fused fibre bundles (known as hexabundles; \citealt{Bland-Hawthorn:2011,Bryant:2014}) with a high (75\%) fill factor. Each bundle contains 61 fibres of $1\farcs6$ diameter resulting in each IFU having a diameter of 15 arcsec. The IFUs, as well as 26 sky fibres, are plugged within the field of view into pre-drilled plates using magnetic connectors. SAMI fibres are fed to the double-beam AAOmega spectrograph \citep{Sharp:2006}. AAOmega allows a range of different resolutions and wavelength ranges. SAMI Galaxy survey observations use the 570V grating, providing coverage between 3700-5700\AA{} (the "blue arm") and the R1000 grating giving coverage between 6250-7350\AA{} (the "red arm"). The spectral resolutions are $R\approx1810$ at 4800\AA{} in the blue arm and $R\approx4260$ at 6850\AA{} in the red arm.

The SAMI DR3 contains fully reduced datacubes of 3068 unique galaxies. Of these, 888 are from the SAMI cluster survey \citep{Owers:2017} and reside in 8 low-redshift massive galaxy clusters. Galaxies span a redshift range of $0.004 <z< 0.095$ and stellar mass range of $\sim10^8-10^{12}$\Msun. Full details of the sample can be found in \citet{Croom:2012} and \citet{Bryant:2015}.

\subsection{Kinematic Morphology Classifications}
\label{sec:kinematic_measurements}

The resolved measurements of the line-of-sight velocity and velocity dispersion of the SAMI sample are described in \citet{vandeSande:2017}. We briefly summarise them here. The SAMI red and blue arm data are combined together by convolving the red arm data to the spectral resolution of the blue. The penalised pixel-fitting code \textsc{pPXF} \citep{Cappellari:2004,Cappellari:2017} is then used to measure the line-of-sight velocity distribution (LOSVD) of each spectrum, parameterised using a Gauss-Hermite series expansion. From the SAMI annular binned spectra, a set of radially-varying optimal templates are found by combining spectra from the MILES stellar library \citep{Sanchez-Blazquez:2006, Falcon-Barroso:2011}. For the spectra in individual spaxels, \textsc{pPXF} is constrained to select from the optimal template for its annulus as well as the spectra from neighbouring annuli. Restricting the available templates in this way was used to prevent template mismatch in low signal-to-noise spaxels. Uncertainties on the LOSVD parameters were estimated by taking the best-fit template, adding appropriate noise and re-fitting 150 times.

\citet{vandeSande:2017} then measure the observed spin-parameter proxy $\lambda_{r}$ for each galaxy, including a seeing correction from \citet{Harborne:2020} and an aperture correction, before deriving an \textit{intrinsic} spin parameter measurement for each object. This is the $\lambda_{r}$ value one would measure if the galaxy was observed edge-on, and is derived assuming that galaxies are simple rotating oblate axisymmetric spheroids which have both varying intrinsic shape and mild anisotropy \citep[see e.g.][]{Cappellari:2007}. Kinematics were measured for 1833 galaxies, but the low completeness below a stellar mass of \lmstar=9.5 led \citet{vandeSande:2017} to remove all galaxies below this mass. The SAMI "kinematic sample" therefore contains 1764 galaxies.

The half-light radius ($r_e$) and $r$-band ellipticity of each galaxy are measured in \citet{D'Eugenio:2021} using the multi-gaussian expansion (MGE) method \citep{Emsellem:1994,Cappellari:2002} on images from GAMA-SDSS \citep{Driver:2011}, SDSS \citep{York:2000}, and VST/ATLAS \citep{Shanks:2013,Owers:2017}.

Finally, galaxies are assigned a kinematic morphological classification based on these value of intrinsic $\lambda_{r}$ and $\epsilon$. We use the boundary in $\lambda_{r}$-- $\epsilon$ space described by \citet{Cappellari:2016}, but our conclusions are unchanged if we instead use the boundary described in \citet{vandeSande:2021a}.

\subsection{Ancillary Measurements}
\label{sec:ancillary}

Stellar masses are derived using aperture-matched $g$ and $i$ band photometry from the GAMA survey \citep{Hill:2011, Liske:2015} for the main SAMI sample and a combination of SDSS Data Release 9 \citep{Ahn:2012} and VST/ATLAS imaging \citep{Shanks:2013,Owers:2017} for the cluster sample. The absolute $i$-band magnitude and the $g-i$ colour of each galaxy are used to derive a stellar mass using the method of \citet{Taylor:2011}; see \citet{Bryant:2015} for details.

The local environment of each object is quantified by measuring the number density of other galaxies surrounding it, $\Sigma$, using the method of \citet{Brough:2013} and \citet{Brough:2017}. $\Sigma_{\mathrm{N,Vlim,Mlim}}$ is defined as the surface density one would derive using the comoving distance to the N\textsuperscript{th} nearest neighbour, within a velocity range of $\pm$V\textsubscript{lim} kms$^{-1}$. Only galaxies with an absolute $r$-band magnitude below $M_r < M_{\mathrm{lim}} - Q$ are included in the calculation. Following \citet{Brough:2017} and \citet{vandeSande:2021} we use $Q = 1.03$, which is
defined as the expected redshift evolution of $M_r$ \citep{Loveday:2015}. This work uses N$=5$, V\textsubscript{lim}$=1000$ kms$^{-1}$ and M\textsubscript{lim}$=-18.6$ mag for galaxies in the SAMI cluster survey, and M\textsubscript{lim}$=-19$ mag for galaxies in the GAMA fields (see \citealt{Croom:2021} for further details). Going forward, we call this measurement \logenv{}.

Star-formation rates (SFRs) are obtained from the spectral energy distribution (SED) fitting code \textsc{magphys} \citep{daCunha:2008,Driver:2016} for the GAMA survey \citep{Gunawardhana:2013,Davies:2016,Driver:2018}. These measurements are estimated by fitting the SED of each galaxy measured in 21 different spectral bands across the UV, optical and far-infrared regions of the electromagnetic spectrum. The \textsc{magphys} template library includes dust emission profiles, such that the measured SFRs are corrected for dust emission.

\subsection{Final sample statistics}

We require our final sample of galaxies to have reliable kinematic measurements, as well as estimates of their stellar mass, environmental surface density and H$\alpha$ star-formation rate. This leaves us with 1676 galaxies. 
Of these, assuming the $\lambda_r$--$\epsilon$ boundary of \citet{Cappellari:2016}, 95
are slow rotators. This is an overall fraction of 5.67\%.
Note that this fraction is simply the number of slow rotators in the catalogue divided by the total number of galaxies; it is not a statement about the true slow rotator fraction of a volume limited sample of the Universe. A volume correction would be necessary to calculate that value, in order to account for the SAMI selection function (but see also \citealt{vandeSande:2021a,vandeSande:2021} and the discussion in Section \ref{sec:volume_correction}).

The galaxies in our sample are shown in Figure \ref{fig:sample}. The top panels show the stellar mass, star-formation rate and \logenv{} for each galaxy, with fast-rotators outlined in black and slow rotators outlined in red. Points without an outline do not have reliable kinematic measurements. Histograms along each axis show the marginal distributions for each parameter.

The bottom panels show the distribution of slow rotators (red points) and fast rotators (blue points) as a function of stellar mass, \logenv{}, $r$-band half-light radius, SFR, ellipticity, mass-weighted metallicity, mass-weighted age and spectroscopic redshift. We plot fast rotators at a $y$ value of 0, and slow rotators at a $y$ value of 1; we then add a random jitter to each point to avoid them being plotted on top of one another. The large points with error bars show the fraction of slow rotators in each bin. We can see that there is a trend for the fraction of slow rotators to increase with increasing stellar mass and also with increasing \logenv{}. However, as we discuss in Section \ref{sec:results}, this is not evidence that environmental surface density positively correlates with being a slow rotator (due to the well-known correlations between environment and other parameters).

\section{The logistic regression model}
\label{sec:logistic_regression}

In this Section, we give a brief introduction to logistic regression before describing in detail the models we fit to our data and the way we compare their performance.

\subsection{An overview of logistic regression}
\label{sec:logistic_regression_introduction}
Logistic regression is a generalisation of standard  linear regression, and is part of a family of statistical tools known as "generalised linear models" (GLMs). It is appropriate when modelling binary response variables: in our case, the quantity we want to model, $y$, is a vector where individual elements $y_i$ take the value 1 if a galaxy is a slow rotator and the value 0 if it is a fast rotator.

Both linear regression and GLMs use a linear combination of predictors (usually written $\bm{X}$) and a vector of model coefficients (usually labelled $\bm{\beta}$) to model the change in a response variable $y$. We discuss our choice of predictors in Section \ref{sec:model_comparison}. In contrast to linear regression, however, GLMs also require a choice of "link function" which maps the linear combination $\bm{X}\cdot\bm{\beta}$ onto a more appropriate domain for the problem at hand. In the case of logistic regression, we model the \textit{probability} that $y_i=1$ as follows:

\begin{equation}
    \label{eqtn:logistic_regression_prob}
    \mathrm{Pr}(y_i=1|X_i=x_i)=p_i=\sigma(X^{\top}_i)\beta,
\end{equation}

where $\sigma(\cdot)$ is the sigmoid function defined as

\begin{equation*}
    \sigma(x)=\frac{1}{1+e^{-x}}
\end{equation*}

This function maps the range (-$\infty$, $\infty$) to (0,1), as is appropriate for modelling probabilities.

We now model the observed $y$ values as being independent and identically distributed, each drawn from a Bernoulli distribution\footnote{i.e. a Binomial distribution with $n=1$ trials} with probability of success given by Equation \ref{eqtn:logistic_regression_prob}. This gives us the following log likelihood function for the $N$ observed data points $y_i$, where the subscript $i$ runs over the number of galaxies in our sample:

\begin{equation}
    \label{eqtn:likelihood}
    \log \mathcal{L} = \sum_{i=1}^{N} y_i \log p_i + (1 - y_i) \log (1 - p_i)
\end{equation}

There are a number of ways to optimise equations \ref{eqtn:logistic_regression_prob} and \ref{eqtn:likelihood} to infer the best fit values of $\bm{\beta}$. We choose to take a Bayesian approach, placing priors on the model parameters and using the probabilistic programming language \texttt{Stan} \citep{Stan} to estimate the resulting posterior distribution.

Finally, we finish this introductory section with a note regarding interpretation of model coefficients. The nonlinearity of the logistic function means that care must be taken when doing so. In particular, unlike in linear regression, the value of the predictor variables at which one wants to evaluate a change in probability of being a slow rotator becomes important. For example, what is the increase or decrease in the probability of being a slow rotator when increasing stellar mass by one dex? The answer changes depending whether you want to know about an increase from $10^{9}\,\mathrm{M}_{\odot}$ to $10^{10}\,\mathrm{M}_{\odot}$ or $10^{10}\,\mathrm{M}_{\odot}$ to $10^{11}\,\mathrm{M}_{\odot}$. The correct way to calculate the change in probability when comparing galaxies with vectors of predictors $\bm{X_1}$ and $\bm{X_2}$ is to calculate:

\begin{equation}
    \label{eqtn:delta_prob}
    \Delta p = \sigma(\bm{\beta}\cdot\bm{X_2}) - \sigma(\bm{\beta}\cdot\bm{X_1})
\end{equation}

\subsection{Model comparison}
\label{sec:model_comparison}

We now describe the steps we take to fit logistic regression models in this work, the details of the different models we investigate and our method for comparing their performance. Our goal is to find a combination of variables which are best able to predict the kinematic morphology of a galaxy. During the following analysis, we centre the values of each of our predictor variables around zero by subtracting their mean. Note that we do not include uncertainties on the observed variables in this modelling process.

For the models presented in this paper we use the \texttt{CmdStanPy} interface\footnote{\url{https://mc-stan.org/cmdstanpy/}} to \texttt{Stan} to perform Hamiltonian Monte Carlo using the No U-Turn sampler \citep{NUTS}. In all cases, we use 4 chains which each take 1000 warmup transitions and 1000 sampling transitions. We ensure that there are no divergent transitions during the fitting and that the Gelman-Rubin convergence statistic $\hat{R}$  \citep{Gelman_Rubin:1992, Vehtari:2021} is equal to 1 for all parameters.

We place the same weakly informative priors on each parameter apart from the intercept term; a Normal distribution centred on zero with standard deviation of 5. In logistic regression, the intercept term is related to the overall fraction of times $y$ takes the value 1. In this case, we have 95
slow rotators in a sample of 1676
galaxies, which is a rate of 5.67\%.
This implies that, if the coefficient of all other terms were zero, the intercept would take the value $\mathrm{logit}(95/1676)=-2.81$, where $\mathrm{logit}=\log(\frac{p}{1-p})$.
We therefore use a Normal distribution centred on -3 with a standard deviation of 5 for the prior on the intercept term, to reflect the fact that we expect its derived value to be large and negative. Reasonable changes to these priors have no effect on our conclusions.

For each of the models described below, we use the method of leave-one-out cross validation (LOO-CV) to derive an "effective log predictive density" score (ELPD$_{\mathrm{loo}}$). This quantity assesses the utility of a model by estimating how well every subset of $N-1$ data points is able to predict value of the $N$\textsuperscript{th}. The ELPD is an alternative to other information criteria such as the Akaike information criterion (AIC: \citealt{AIC}) or Bayesian information criterion (BIC: \citealt{BIC}), which we also calculate for our models.

We use the package \texttt{arviz}\footnote{\url{https://arviz-devs.github.io/arviz/}} \citep{arviz:2019} to perform Pareto-smoothed importance sampling leave-one-out cross-validation \citep[PSIS-LOO CV;][]{Vehtari:2017}, which exploits the conditional independence of the observations to derive an ELPD$_{\mathrm{loo}}$ score for the model without having to refit the model $N$ times. The ELPD$_{\mathrm{loo}}$ score for each model is defined as

\begin{equation}
    \mathrm{ELPD_{loo}} = \sum_{i=1}^{i=N} \log p(y_i|y_{-i})
\end{equation}

\noindent where
\begin{equation}
    \log p(y_i|y_{-i}) = \int p(y_i|\bm{\beta})p(\bm{\beta}|y_{-i})d\bm{\beta}
\end{equation}

\noindent is the leave-one-out predictive density given the data \textit{without} the $i$th data point. In practice, this integral is approximated using samples from the posterior provided by \texttt{Stan} (see Section \ref{sec:logistic_regression_introduction}).

Note that we do not use the classification accuracy of each model in assessing its utility. It would be possible to make a binary prediction for a given galaxy (if, say, the derived probability of it being a slow rotator was greater than some threshold) and then compare these LOO-CV predictions to their true classification. Doing so would ignore the additional information which is captured in the derived $p_i$ values, however. Classification accuracy can also be a potentially misleading scoring metric. In our sample of 1676
galaxies with 95 slow rotators, a model which classified all galaxies as fast rotators would result in $\approx 95\%$ accuracy yet would clearly be of limited use.

\section{Results}
\label{sec:results}

We now discuss the details of the logistic regression models we build, beginning with the simplest and then including further variables of interest.

\subsubsection{Mass and Environment}

Firstly, we investigate a model which uses the stellar mass and environmental surface density of each galaxy. This is a natural starting point for our investigation, building upon previous studies who find that these quantities correlate with kinematic morphology \citep[e.g][]{Cappellari:2011b, Cappellari:2016, vandeSande:2021}.

When we do so, we find that the coefficient of \lmstar{} is $3.073\pm0.320$, the coefficient of of \logenv{} is $0.334\pm0.145$ and the intercept is $-4.606\pm0.262$. This shows that, in agreement with \citet{vandeSande:2021}, at fixed mass there is a residual dependence of kinematic morphology on local environment \textit{when one only considers these two quantities}.

\subsubsection{Further predictors and our best-fit model}
\label{sec:bestfitpredictors}

Compared to subdividing a sample into separate bins at fixed quantities, logistic regression allows us to include further quantities in our model without running into the issue of small number statistics. We therefore investigate the inclusion of other variables and assess their utility in the fit.

We choose to study the influence of stellar mass, environment, $r$-band half-light radius ($r_e$), H$\alpha$ star-formation rate, redshift and two binary variables: a "quenched flag" which takes the value 1 if a galaxy is quenched and zero otherwise, and an "ellipticity flag" which takes the value 1 if a galaxy has $\epsilon<0.4$ and 0 otherwise. We choose these quantities because they are conceivably of use for selecting slow rotators: slow rotators tend to be objects which have quenched their star-formation, lie on a well-defined region of the mass-size plane and are always rounder than $\epsilon =0.4$ \citep{Cappellari:2016}. This gives us 7 predictors. We note that these quantities can be measured from single-fibre spectroscopic surveys and/or wide-field imaging. This is a deliberate choice, as we aim to be able to predict the expensive measurement of a galaxy's kinematic morphology without including data that requires spatially-resolved spectroscopy.

In each case, we will use a model's AIC, BIC and ELPD$_{\mathrm{loo}}$ score to asses its utility. We perform "best-subset regression" \citep[e.g.][]{Hocking:1967} using an exhaustive search, meaning we fit $2^7=128$ different models corresponding to all linear combinations of our predictors (excluding interaction terms). The best AIC, BIC and ELPD$_{\mathrm{loo}}$ scores for models varying number of predictors are shown in Figure \ref{fig:model_comparison}.

The lowest AIC and ELPD$_{\mathrm{loo}}$ scores occur for the same model: $y \sim \log_{10}(\mathrm{M}_* / \mathrm{M}_{\odot}) + \log_{10}(r_e / \mathrm{kpc}) + \log_{10}(\mathrm{SFR}/ \mathrm{M}_{\odot} \mathrm{yr}^{-1}) + \epsilon_{0,1}$. The BIC prefers a more parsimonious model, removing the dependence on $r_e$: $y \sim \log_{10}(\mathrm{M}_* / \mathrm{M}_{\odot}) + \log_{10}(\mathrm{SFR}/ \mathrm{M}_{\odot} \mathrm{yr}^{-1}) + \epsilon_{0,1}$. This is not unexpected: BIC is generally known to prefer models with fewer predictors\footnote{For $n$ data points and $k$ parameters in the model, BIC penalises the likelihood by a factor of $k\log(n)$, whereas the AIC penalty term is $2k$.}.

We choose to use $y \sim \log_{10}(\mathrm{M}_* / \mathrm{M}_{\odot}) + \log_{10}(r_e / \mathrm{kpc}) + \log_{10}(\mathrm{SFR}/ \mathrm{M}_{\odot} \mathrm{yr}^{-1}) + \epsilon_{0,1}$ as our preferred model going forward, since it is selected by two out of three information criteria. The best-fitting values of each parameter in this logistic regression model are shown in Table \ref{tbl:best_model}.

\begin{table}
\centering
\caption{The best-fit parameters in our preferred model with predictors \lmstar, \logsfr{}, \logre{} and a binary flag if $\epsilon$ is less than 0.4, as described in Section \ref{sec:bestfitpredictors}. We show the posterior mean, the standard deviation and the 2.5th, 16th, 50th, 84th and 97.5th percentiles of each parameter.}
\label{tbl:best_model}
\begin{tabular}{lrrrrrrr}
\toprule
 & Mean & $\sigma$ & $2.5^{\mathrm{th}}$ & $16^{\mathrm{th}}$ & $50^{\mathrm{th}}$ & $84^{\mathrm{th}}$ & $97.5^{\mathrm{th}}$ \\
\midrule
$\beta_{\mathrm{M}_{*}}$ & 2.97 & 0.39 & 2.23 & 2.59 & 2.97 & 3.36 & 3.77 \\
$\beta_{\mathrm{SFR}}$ & -0.88 & 0.17 & -1.23 & -1.06 & -0.88 & -0.71 & -0.55 \\
$\beta_{\mathrm{r_e}}$ & 1.09 & 0.65 & -0.19 & 0.44 & 1.09 & 1.73 & 2.37 \\
$\beta_{\epsilon}$ & 4.97 & 1.61 & 2.52 & 3.44 & 4.73 & 6.51 & 8.77 \\
$c$ & -11.23 & 1.69 & -15.12 & -12.87 & -11.03 & -9.61 & -8.52 \\
\bottomrule
\end{tabular}
\end{table}

We also investigate regularisation as an alternative to our best-subset regression approach, finding very similar results. This is discussed in Appendix \ref{sec:appendix}.

\subsubsection{Stellar population parameters}

In addition to these quantities, we also investigate the utility of including predictors related to a galaxy's stellar population, namely its mass-weighted age and mass-weighted stellar metallicity derived from full-spectral fitting. We separate this investigation into a separate subsection because such stellar population measurements are not as widely available as the variables discussed in \ref{sec:bestfitpredictors}, as well as being more dependent on the precise details of the modelling approach taken during their calculation.

The mass-weighted age and stellar metallicity for each galaxy are measured from a spectrum extracted within one effective radius. We use the code \textsc{pPXF} method \citep{Cappellari:2004,Cappellari:2017} combined with the MILES simple stellar population templates \citep{Vazdekis:2015}. Full details can be found in \citet{Vaughan:2022}

We perform a second exhaustive search with 9 parameters: the 7 from \ref{sec:bestfitpredictors} plus age and [Z/H]. We find that the model with the lowest ELPD$_{\mathrm{loo}}$ contains the variables \lmstar, \logre, \logsfr, $\epsilon_{0,1}$ and [Z/H]. However, the difference in ELPD$_{\mathrm{loo}}$ between this model and the best performing model from Section \ref{sec:bestfitpredictors} is only $2.2\pm4.2$, i.e the models with and without [Z/H] are not significantly different. This is also true for a number of other models: the set of predictors \lmstar, \logre, \logsfr, $\epsilon_{0,1}$, [Z/H] and age also has an ELPD$_{\mathrm{loo}}$ within 1$\sigma$ of the best performing model, as do the models which contain \lmstar, \logsfr, $\epsilon_{0,1}$ and either age, [Z/H] or both at the same time.

What we find, therefore, is that a number of different combinations of variables give results which are indistinguishable according to our scoring criteria. Going forward, we choose to continue using the best-fit model from Section \ref{sec:bestfitpredictors} (i.e. the model without [Z/H] or stellar age) since it contains variables which are more widely measured in large spectroscopic surveys.

\subsection{Model Checking}

\begin{figure*}
    \includegraphics[width=\textwidth]{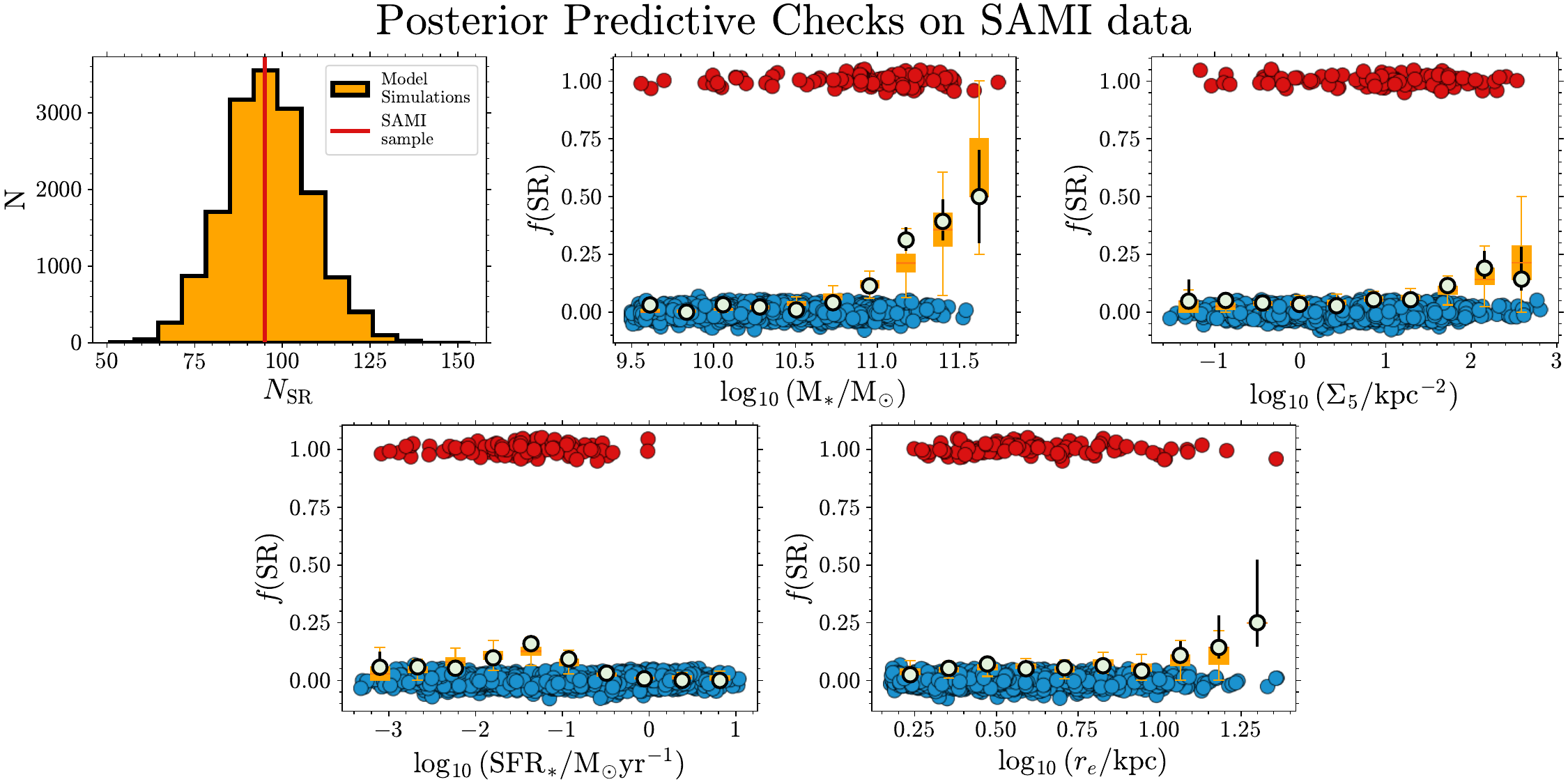}
    \caption{Posterior predictive checks to examine the performance of the best performing model using the variables discussed in \ref{sec:bestfitpredictors}. We use the observations of stellar mass, star-formation rate and a binary flag if a galaxy has $\epsilon<0.4$ as predictors and simulate 4000 possible sets of observations. Each of these simulations is a series of slow rotator/fast rotator classifications for the galaxies in our sample. We then compare these simulated datasets to our observed SAMI kinematic morphologies. In each panel, results from these simulations are shown in orange. \textbf{Top left}: A histogram of the number of slow rotators in the simulations, with the observed number of slow rotators in the SAMI data shown in red. \textbf{Remaining panels}: The slow rotators in our SAMI data and simulations as a function of stellar mass, environmental surface density, star-formation rate  and half-light radius (bottom right). The large circular points with error bars show the fraction of SRs in the SAMI data in bins of the $x$-axis variable. The orange box-and-whisker plots show the same quantity from the simulated datasets, which are consistent with the observations. We note that our model recovers the increase in the slow-rotator fraction towards denser environments (top right panel) despite the fact that environmental surface density is not included in the model and does not show a statistically significant correlation with $f(SR)$ at fixed mass and star-formation rate.}
    \label{fig:PPCs}
\end{figure*}

An important step in any modelling workflow is to assess the model's performance \citep[e.g.][]{Gelman:2020}. In Figure \ref{fig:PPCs}, we show posterior predictive checks for the best-fitting model of Section \ref{sec:bestfitpredictors} compared with the observed SAMI data.

Posterior predictive checking uses draws from the model posterior to create simulated observations which are then compared to the observed data. For a well specified and appropriate model, these quantities should appear indistinguishable \citep{Rubin:1984,Gelman:1996}.

The first panel of Figure \ref{fig:PPCs} shows a histogram of the number of slow rotators in 4000 posterior predictive simulations from our model. The true value in the SAMI data is shown in red.

The remaining panels of Figure \ref{fig:PPCs} recreate the bottom panels of Figure \ref{fig:sample}, showing the fraction of slow rotators in our sample as a function of stellar mass, environment and star-formation rate. Model simulations of the slow-rotator fraction in the same bins are shown as orange box-and-whisker plots. These give the median, 16\textsuperscript{th} to 84\textsuperscript{th} percentile range and minimum and maximum values of $f(\mathrm{SR})$ in each bin. The observations are in excellent agreement with the model simulations. We also note that the model recovers the observed correlation between $f(\mathrm{SR})$ and \logenv{} despite not containing any dependence on environment.

\begin{figure}
    \centering
    \includegraphics[width=0.5\textwidth]{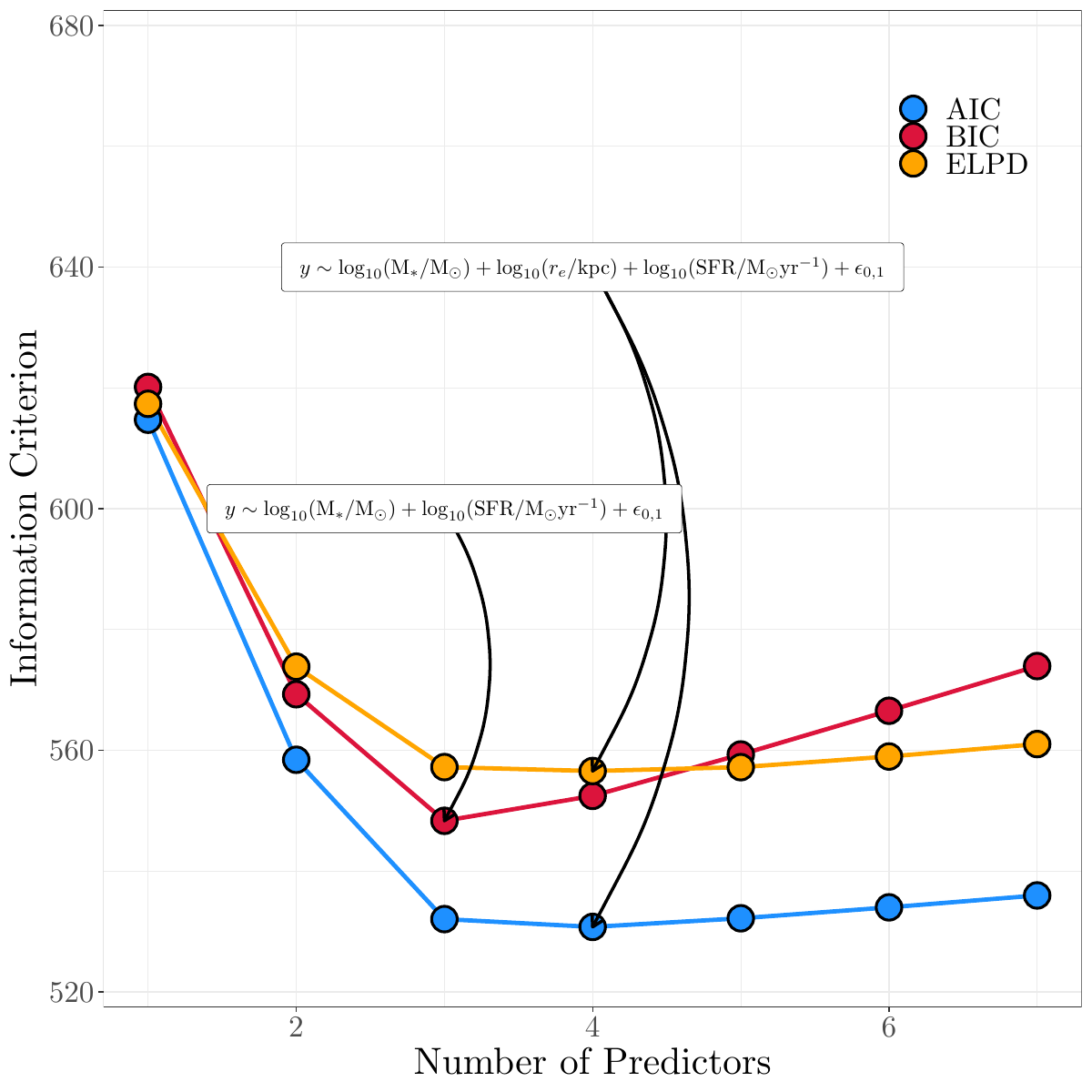}
    \caption{A comparison of the best AIC, BIC and ELPD$_{\mathrm{loo}}$ scores for models with varying numbers of predictors. The overall best models (those with lowest information criteria scores) are highlighted. Note that we have converted the ELPD$_{\mathrm{loo}}$ scores to information criteria by multiplying by $-2$ (i.e. to the deviance scale).}
    \label{fig:model_comparison}
\end{figure}

\subsection{Model Validation using MaNGA}

The Mapping Nearby Galaxies at Apache Point Observatory (MaNGA) survey \citep{Bundy:2015} has observed over 10,000 nearby galaxies with a multiplexed integral-field spectrograph, resulting in a sample of kinematic observations comparable to the SAMI observations used in this work (and described in Section \ref{sec:kinematic_measurements}). Observations from the MaNGA survey therefore make an ideal "test set" for our logistic regression model.

We use the MaNGA kinematic measurements from \citet{Fraser-McKelvie:2022}, which provide measurements of $\lambda_r$ and $\epsilon$ for 4742 nearby galaxies. Each of these galaxies also has stellar mass and star-formation rate measurements from the GALEX-SDSS-WISE Legacy catalogue (GSWLC; \citealt{Salim:2016,Salim:2018}) and half-light radii measured using an elliptical Petrosian model \citep{Blanton:2011}.

Note that the MaNGA $\lambda_r$ measurements we use have been corrected for beam-smearing using the empirical relation of \citet{Harborne:2020} but have \textit{not} been deprojected (such that the velocity values are those which would be measured if the galaxy was to be observed edge-on).

Using the kinematic boundary of \citet{Cappellari:2016}, 691 MaNGA galaxies with \lmstar$>9$ are observed to be slow rotators, an overall fraction of 14.6\% (691 / 4742). This fraction is larger than the 10.1\% of \lmstar$>9$ galaxies in the catalogue of \citet{Graham:2018} using a preliminary MaNGA data release (although it should be noted that the beam-smearing correction techniques used by \citet{Graham:2018} and \citet{Fraser-McKelvie:2022} are different: see \citealt{Harborne:2020}). Note that neither of these numbers are representative of the fraction of slow rotators in a volume limited sample of the Universe; the MaNGA selection function is (by design) \textit{not} a representative sample of nearby galaxies (see e.g. \citealt{Bundy:2015}). \citet{Fraser-McKelvie:2022} correct for this effect by applying a volume correction to their catalogue, concluding that their observations represent a slow rotator fraction of 6\% in the nearby Universe (in the mass range $9.75<$\lmstar$<11.75$).

Using the stellar masses, sizes, star-formation rates and ellipticities of the \citet{Fraser-McKelvie:2022} sample, the logistic regression model of Section \ref{sec:logistic_regression} predicts a SR fraction of $12.82\pm1.49$

in this sample. This is only $\approx1.2\sigma$ away from the observed value of 14.6\%, highlighting the effectiveness of the model at making predictions on unseen data.

\begin{figure*}
    \includegraphics[width=0.95\textwidth]{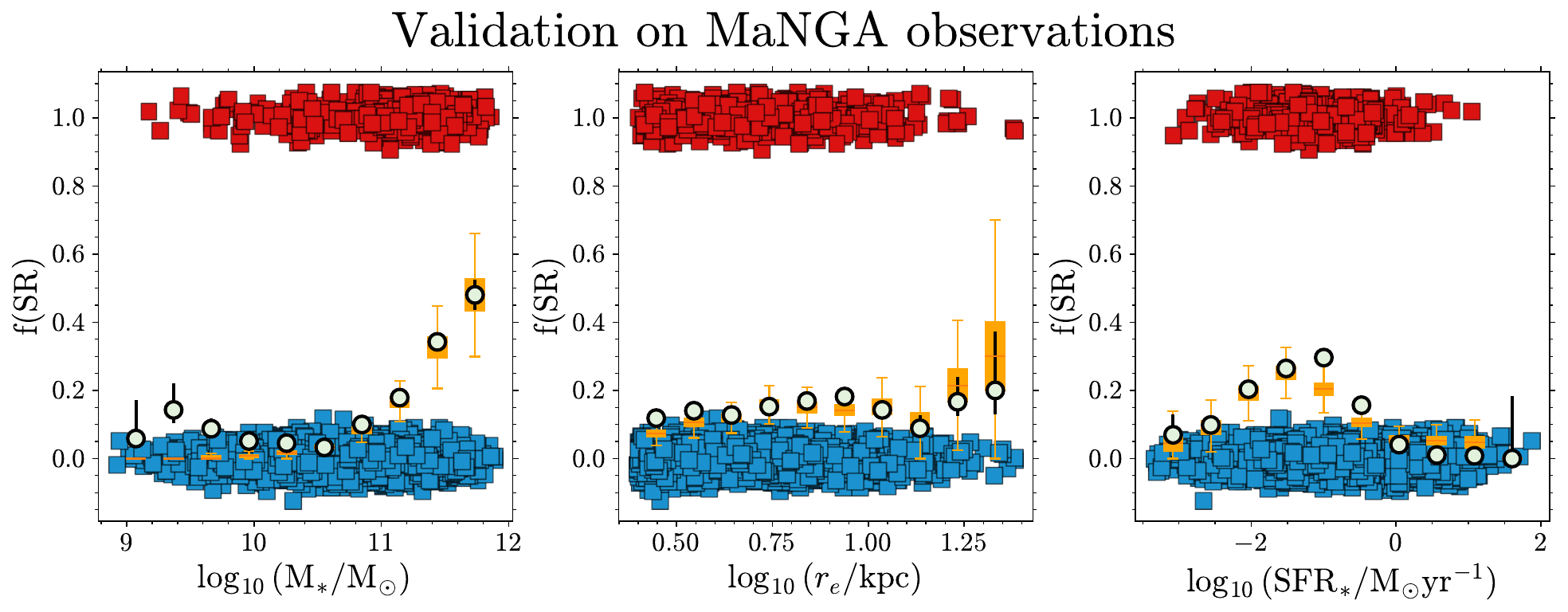}
    \caption{Slow rotators in bins of stellar mass (left), half-light radius (centre) and star-formation rate (right) from MaNGA observations of 4742 nearby galaxies \protect\citep{Fraser-McKelvie:2022}. Similarly to Figure \ref{fig:PPCs}, the large circular points with error bars show the fraction of SRs in bins of the $x$-axis variable. The orange box-and-whisker plots show the same SR fractions predicted from our logistic regression model. We see an excellent agreement between our model predictions and the MaNGA observations, which shows the effectiveness of our logistic regression model at making predictions on unseen data.}
    \label{fig:manga_predictions}
\end{figure*}

As well as closely matching the overall fraction of slow rotators in the MaNGA data, Figure \ref{fig:manga_predictions} also highlights the performance of the model at predicting the fraction of slow rotators in bins of stellar mass, half-light radius and star-formation rate. We see excellent agreement between the data and model across nearly all bins. The only notable deviations occur at stellar masses less than $10^{10}$ \Msun, where the MaNGA observations show a small increase in SR fraction which is not observed in the SAMI observations (although note that the MaNGA sample contains galaxies down to $10^{9}$ \Msun whereas the SAMI sample excludes objects below $10^{9.5}$ \Msun). It is these low-mass galaxies which fall into the slow rotator selection box which are missed by our model, leading to the underprediction of the overall fraction of slow rotators in the full MaNGA sample.

We note that there are many reasons which may cause low mass galaxies to have low $\lambda_r$ values, of which being a true slow rotator is only one. For example, \citet{Fraser-McKelvie:2022} describe that, of 14 star-forming slow rotators below \lmstar$=10.75$ in their sample, only one shows a high stellar velocity dispersion for its mass with little rotational support. Instead, these objects tend to either be undergoing mergers, contain kinematically decoupled cores and other kinematic features or are simply viewed almost exactly face on (and therefore have very small line-of-sight velocities).

Recent work has hinted at a decrease in $\lambda_r$ with decreasing masses below $\approx 10^{9.5}$\Msun \citep{FalconBarroso:2019, Scott:2020,vandeSande:2021a}, however, such that some low-mass galaxies truly do fall into the slow rotator regime. Confirming this result requires a larger sample size of high spectral-resolution IFU observations of low-mass galaxies\footnote{e.g. see the forthcoming IFU survey Hector \citep{Bryant:2020}}, however, and including this effect in our model is beyond the scope of the current work. We therefore instead simply caution against using this model on galaxies with a stellar mass below $10^{9.5}$\Msun (i.e. the lower mass limit of the SAMI kinematic sample).

\section{Discussion}
\label{sec:discussion}

The main conclusion of this work is that knowing a galaxy's observed environment (as measured through its value of $\Sigma_5$) gives you no further information about whether it is a slow rotator once stellar mass, size, star-formation rate and ellipticity are known.

This result may be surprising and counter-intuitive, since slow rotators are preferentially found in dense environments \citep[e.g][]{Cappellari:2011b, Houghton:2013, D'Eugenio:2013, Scott:2014, Fogarty:2014}. We can see this in our own data: in Figure \ref{fig:sample}, the fraction of slow rotators increases towards larger values of \logenv{}.

However, the key insight offered by our analysis is the ability to study the effect of each parameter \textit{whilst keeping all others fixed}. Plots like Figure \ref{fig:sample}, and those shown in numerous previous studies, are potentially misleading as \logenv{} is also correlated with \lmstar{}, \logsfr{} and many other parameters. Slow rotators are preferentially found in dense environments, therefore, not because their current environmental surface density affects their kinematic morphology but because that is where massive quenched galaxies tend to reside.

This conclusion is in agreement with \citet{Veale:2017}, \citet{Greene:2017} and \citet{Brough:2017}, who find that the kinematic morphology-density relation can be accounted for by the well-known correlation between mass and environment. Our work extends this result to quantify which other parameters add useful information for identifying slow rotators (namely their star-formation rate, half-light radius and their observed ellipticity).

Our findings contribute to the body of work which shows that the effects of environment on galaxy properties are subtle once stellar mass is controlled for, particularly for high-mass objects \citep[e.g.][]{Blanton:2009}. \citet{Bamford:2009} conclude that galaxy morphology displays only a weak environmental trend at fixed stellar mass (see also \citealt{Alpaslan:2015}), whilst \citet{Thomas:2010} show the same is true for a number of stellar population scaling relations. \citet{Goddard:2017}, \citet{Zheng:2017} and \citet{Ferreras:2019} also find that stellar population \textit{gradients} are only a very weak function of local environment at fixed mass (and see also \citealt{Santucci:2020} who find similar gradients between satellites and centrals). A lack of environmental dependence has also been shown for the star-formation rates of star-forming galaxies \citep{Wijesinghe:2012, Muzzin:2012}, as well as the spatial distribution of star-formation within star-forming objects \citep{Spindler:2018}.

The strongest correlation in our model is between kinematic morphology and stellar mass. Stellar mass (or absolute magnitude) has long been known to correlate with kinematic quantities \citep[e.g.][]{Davies:1983, Bender:1992, Emsellem:2007}, with some authors proposing a critical mass of \lmstar=11.3 below which galaxies cannot be classified as "true" slow rotators (see e.g. the discussion in \citealt{Cappellari:2016} and \citealt{Graham:2019b}).

Figure \ref{fig:sample} shows that the fraction of slow rotator galaxies in the SAMI sample increases sharply slightly below this critical mass (although it should also be noted that 86\% of the slow rotators in the survey have masses below \lmstar=11.3). Whilst not containing an explicit "critical value", our model displays behaviour which is compatible with this idea. By definition, logistic regression relies on non-linear behaviour to model the probability of "success" as a function of the input parameters, as shown in Equation \ref{eqtn:logistic_regression_prob}. This nonlinearity means that a small difference in stellar mass reflects a larger increase in slow rotator probability at larger stellar masses than small ones. Quantitatively, for two otherwise-identical galaxies (with with \logsfr=-2, \logre=0.5 and ellipticity less than 0.4), a galaxy with \lmstar=11.4 is 6\% more likely to be a SR than one with \lmstar=11.3. On the other hand, the difference in SR probability is only 0.01\% if the two stellar masses are \lmstar=10 and \lmstar=10.1. In coarsely binned data, such non-linear behaviour may appear identical to a hard boundary between the two classes.

We see a very strong positive dependence on the $\epsilon$-flag variable, reflecting the fact that there are no slow rotators rounder than $\epsilon=0.4$ by definition (given the \citealt{Cappellari:2016} boundary in the $\lambda-\epsilon$ plane). The model also finds a strong negative correlation with star-formation rate, such that galaxies with smaller star-formation rates are more likely to be SRs. This is due to the fact that almost all slow rotators have quenched their star-formation, with the handful of slow rotators falling on the star-formation rate "main sequence" likely to be kinematically complicated systems. The weakly positive correlation with half-light radius implies that, at fixed mass and star-formation rate, a larger galaxy is more likely to be a slow rotator. This result is slightly counterintuitive, since other properties of massive early-type galaxies such as age, metallicity, velocity dispersion and M/L ratio increase with \textit{decreasing} half-light radius \citep[e.g.][]{Cappellari:2016}. We note, however, that the coefficient of \logre{} in the model is only $\approx$1.7 $\sigma$ away from zero.

\subsection{The fraction of slow rotators in a volume limited sample of the nearby Universe}
\label{sec:volume_correction}

We now highlight another strength of modelling the fraction of slow rotators using logistic regression: the ability to use \textit{post-stratification} to predict the fraction of slow rotators in other galaxy surveys.

In Section \ref{sec:sample_selection}, we discussed the properties of the SAMI sample used to derive the logistic regression model in Section \ref{sec:logistic_regression}. As noted, the SAMI survey is not volume limited (see \citealt{Bryant:2015} and \citealt{vandeSande:2021} for further discussion). Once we have fit the model to the SAMI galaxies, however, we can apply it to any given sample of galaxies and predict the fraction of slow rotators which would be observed. In particular, we can predict the fraction of slow rotators which would be observed in a volume limited sample of the nearby Universe.

In practice, we use the Sloan Digital Sky Survey Data release 17 \citep{SDSS_DR17} and select a random sample of 10,000 galaxies which have measurements of redshift, stellar mass, fibre star-formation rate, half-light radius and and axis ratio ($b/a$) from the tables \texttt{SpecObjAll}, \texttt{PhotoObjAll} and \texttt{galSpecExtra}.

We select galaxies which have a stellar mass greater than $10^{9.5}\mathrm{M}_{\odot}$ and a redshift between 0.01 and 0.1. We also require that they have a value of \texttt{SpecObjAll.class}  which is not "\texttt{STAR}" and a \texttt{SpecObjAll.zWarning} value of 0.

In this sample of SDSS galaxies, we predict a slow-rotator fraction of $3.79\pm0.40$
as compared with a value of 5.67\%
in the SAMI catalogue used in this work. This difference between surveys is to be expected given their different selection functions.

\subsection{The formation of slow rotators}

Recent studies of both observations and simulations suggest that slow rotators undergo two distinct phases in their formation: a period where their star-formation quenches and a period where their kinematics are transformed to become dominated by random motion \citep{Cortese:2019, Park:2022, Lagos:2022}. Using the large volume EAGLE simulation \citep{Schaye:2015}, for example, \citet{Lagos:2022} show that quenching of their slow rotators generally happens $\approx$2 Gyrs before their kinematic transformation.

Whilst there are a number of pathways to kinematic transformation, the most common is thought to be via galaxy-galaxy mergers \citep{Naab:2014, Penoyre:2017, Lagos:2018}. Interestingly, it has been found to be important that the progenitors of today's slow rotators are quenched \textit{before} their kinematic transformation \citep{Lagos:2022}. Quenching is required for galaxy mergers to lower the spin parameter $\lambda_r$ more effectively, since a star-forming galaxy undergoing a major merger can reform its disc after the interaction and remain a fast rotator \citep{Penoyre:2017, Lagos:2017, Lagos:2022}.

As discussed in \citet{Cappellari:2016} and \citet{Brough:2017}, one hypothesis is that today's slow rotators always form at the centres of particularly massive dark-matter halos. This allows them to quickly grow in stellar mass, after which they quench and transform their kinematics by merging with infalling satellite galaxies. This process must happen before their dark matter halo becomes too large; the velocity dispersions of massive clusters prevent satellite galaxies from merging efficiently (e.g. \citealt{Ostriker:1980}).

During this process, their dark-matter halo will either accrete further sub-halos to become a galaxy cluster at $z=0$ or itself be accreted by a larger overdensity. This naturally explains the observation that slow rotators in clusters which are not found at the bottom of the cluster potential tend to be found at locations of substructure within the cluster \citep{Cappellari:2016, Graham:2019}.

If the processes which create slow rotators happens quickly at high redshift, then this picture is also a natural explanation for the lack of dependence of kinematic morphology on a galaxy's local environment at $z=0$. Two identical halos in the early Universe can both lead to the creation of slow rotators at their centre, yet end up residing in regions of the Universe with entirely different values of \logenv{} today if one is accreted into a larger structure whilst the other is not.

\subsection{Future work}

The key variables in our simple logistic regression model turn out to be easily measurable quantities for large imaging or single fibre surveys: stellar mass, half-light radii, star-formation rates and observed ellipticity. In practice, there are a number of other quantities which could be included in the model to make more precise predictions on individual objects.

For example, training a convolutional neural network on image cutouts of galaxies would be an obvious next step. There has already been some work to select SR "candidates" by visual classification from photometry alone \citep{Graham:2019}, and combining our logistic regression model with computer vision techniques could be a fruitful avenue to highlight galaxies in imaging surveys for follow-up observations, or to asses the fraction of slow rotators in a sample without requiring expensive spectroscopic observations. Such models could also be useful for large surveys to predict the number of slow rotators they will observe, or used during a survey's selection phase to target SRs specifically.

We also note that, in the interests of simplicity and to match previous work, we have used parametric cuts in the $\lambda-\epsilon$ plane to select our slow rotator sample. Recent studies have shown, however, that relying on such boundaries for selecting slow rotators can lead to significant contamination \citep{vandeSande:2021a,Lagos:2022}. An alternative approach for selecting a pure sample of SRs is to rely on visual classifications from a group of expert classifiers. The classifications need not be limited to simply slow and fast rotators, however; for example, \citet{vandeSande:2021a} used visual classifications to split their galaxies into "obvious" and "non-obvious" rotators, with each class being further subdivided into galaxies with and without "features" in their kinematic maps (corresponding to kinematically-decoupled cores, two-$\sigma$ galaxies, etc). Multinomial regression, a generalisation of logistic regression to more than two discrete outcomes, would be a natural extension of the current work to such data.

\section{Conclusions}
\label{sec:conclusion}

This work uses data from the SAMI galaxy survey to build a logistic regression model which assesses the correlation between a galaxy's kinematic morphology-- whether it is a fast rotator or a slow rotator-- and a range of other properties.

\begin{itemize}
    \item Our chosen model uses stellar mass, half-light radius, star-formation rate and a binary variable based on whether the galaxy's observed ellipticity is less than 0.4 (see Figure \ref{fig:model_comparison} and Section \ref{sec:model_comparison}).
    \item We find that a galaxy's environment-- measured by its environmental surface density, \logenv{}-- is not a useful parameter in our model. When only considering fixed stellar mass, there is a residual correlation between kinematic morphology and environment. However, when including other properties such as size, star-formation rate and observed ellipticity, this residual correlation disappears.
    \item We therefore conclude that the observed kinematic morphology-density relation can be entirely explained by the well-known correlations between environment and other properties. Whilst slow rotators \textit{are} preferentially found in dense environments (see the bottom panels of Figure \ref{fig:sample}), this is due to the fact that dense environments are more likely to host massive, quenched elliptical galaxies.
    \item We test our model by performing posterior predictive checks, comparing predictions from the model against the data used to find the best-fit parameters (Figure \ref{fig:PPCs}). We also test our model on unseen data, by predicting the slow-rotator fraction in the MaNGA survey as a function of several variables (Figure \ref{fig:manga_predictions}). In both cases, our model recovers the observed trends very accurately.
    \item We applied the model derived from the SAMI Galaxy Survey to a volume-limited sample of galaxies from the Sloan Digital Sky Survey (SDSS) to estimate the slow-rotator fraction of the nearby Universe. Above a stellar mass of $10^{9.5}$\Msun and in a redshift range of 0.01 -- 0.1, we estimate the fraction of slow rotators to be $3.79\pm0.40$
          slightly lower than the 5.67\% found in SAMI.
\end{itemize}

This study adds to the body of work which finds that the influence of environment on galaxy properties is subtle, especially at fixed mass and other quantities. It also highlights the utility and applicability of more generalised statistical techniques (which are common in other fields) to astronomical applications.

\section*{Acknowledgements}

We would like to thank the anonymous referee who's comments improved this work.
The SAMI Galaxy Survey is based on observations made at the Anglo-Australian Telescope. The Sydney-AAO Multi-object Integral field spectrograph (SAMI) was developed jointly by the University of Sydney and the Australian Astronomical Observatory. The SAMI input catalogue is based on data taken from the Sloan Digital Sky Survey, the GAMA Survey and the VST ATLAS Survey. The SAMI Galaxy Survey is supported by the Australian Research Council Centre of Excellence for All Sky Astrophysics in 3 Dimensions (ASTRO 3D), through project number CE170100013, the Australian Research Council Centre of Excellence for All-sky Astrophysics (CAASTRO), through project number CE110001020, and other participating institutions. The SAMI Galaxy Survey website is http://sami-survey.org/.
JvdS acknowledges support of an Australian Research Council Discovery Early Career Research Award (project number DE200100461) funded by the Australian Government. JJB acknowledges support of an Australian Research Council Future Fellowship (FT180100231).

\section*{Data Availability}

The data used in this work are publicly available as part of the SAMI Galaxy Survey Data Release 3 \citep{Croom:2021}. They may be accessed at \url{https://datacentral.org.au/}.



\bibliographystyle{mnras}
\bibliography{bibliography} 




\appendix

\section{Regularised Logistic Regression}
\label{sec:appendix}

In Section \ref{sec:logistic_regression}, we built and analysed our logistic regression model by adding predictors and assessing the change in various information criteria. Another approach is to use a "regularised" regression, where all predictors are included at once but a penalty term $\lambda$ is added to the likelihood to shrink the coefficients of extraneous terms to be small.

\begin{figure*}
    \includegraphics[width=0.8\textwidth]{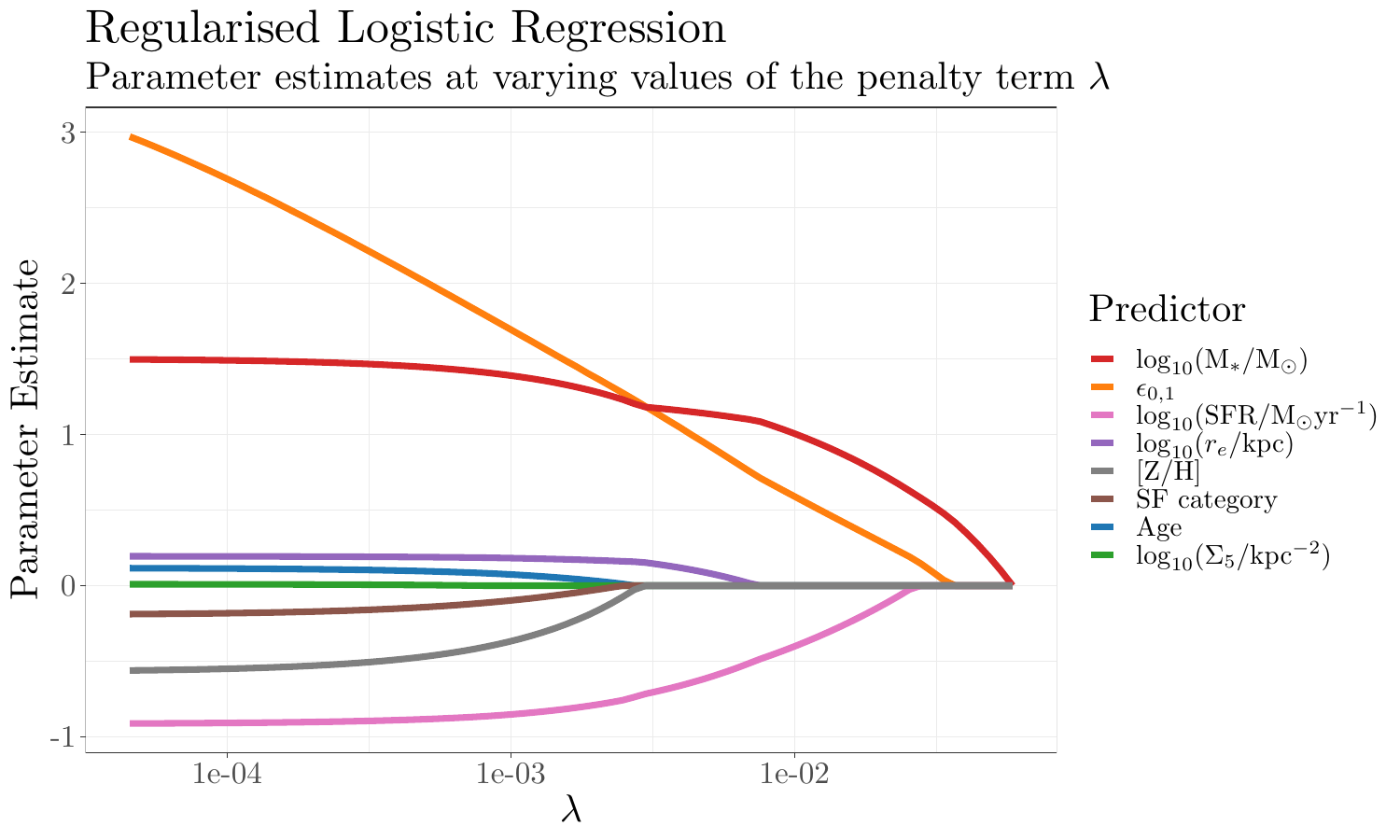}
    \caption{The coefficients of each predictor for varying values of the penalty term $\lambda$. As $\lambda$ increases (to the right), the most important coefficients in the model are the last to be regularised to zero. The legend on the right gives the variables in reverse order of when they drop out of the regression.}
    \label{fig:lasso}
\end{figure*}

The regularised logistic regression implementation is provided by the \textsc{R} package \texttt{glmnet}. We choose to use LASSO regression \citep{LASSO} which uses a penalty term of the form $\lambda \sum_{i=1}^{j}|\beta_j|$, where $\beta_j$ is the coefficient of the $j$\textsuperscript{th} predictor and $\lambda$ is a tunable parameter in the model. Using the same SAMI dataset as Section \ref{sec:logistic_regression}, Figure \ref{fig:lasso} shows the coefficients of each predictor as a function of $\lambda$. As $\lambda$ increases, more and more coefficients tend towards zero. We find that the last four coefficients remaining in the model are \lmstar, $\epsilon_{0,1}$, \logsfr and \logre. These are the same coefficients as used in the model with the best AIC and ELPD\textsubscript{LOO} scores in Section \ref{sec:logistic_regression}.

\begin{figure}
    \includegraphics[width=\columnwidth]{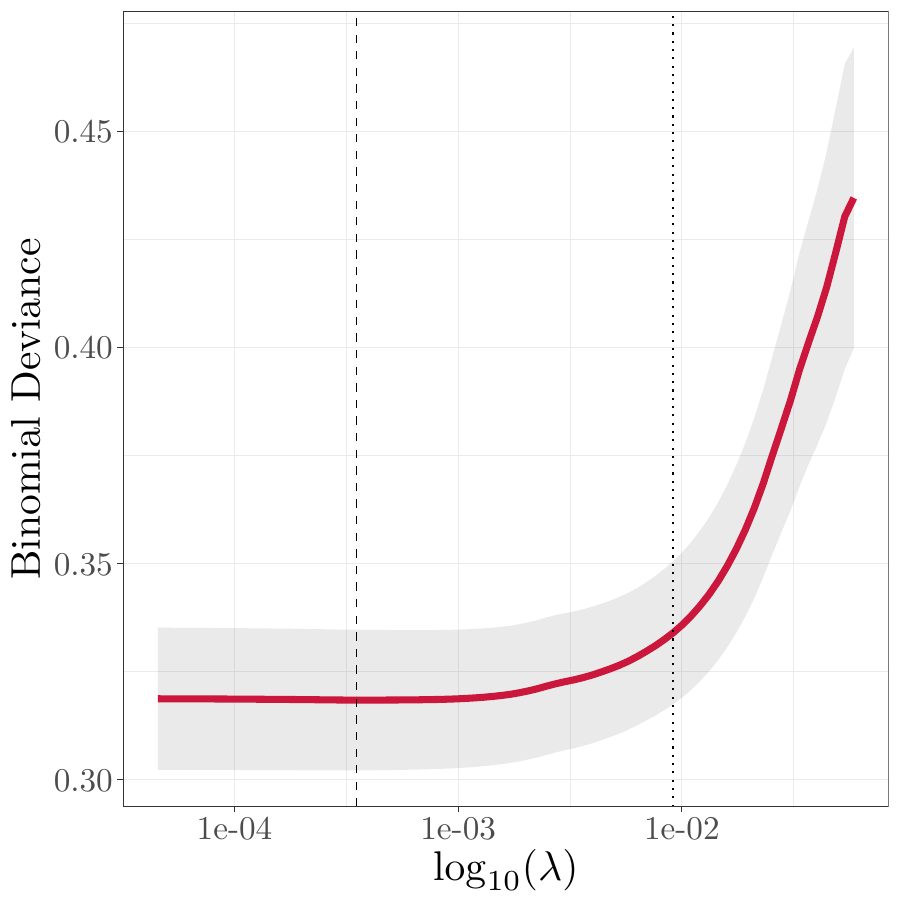}
    \caption{5-fold cross-validation using the SAMI dataset from Section \ref{sec:logistic_regression}. The red-line shows the mean Binomial Deviance, $D$, as a function of $\lambda$, with the shaded regions showing the confidence intervals from the cross-validation. The minimum value of deviance occurs at the dashed line, for a model containing all 8 predictors. We choose to use a value of $\lambda$ where the deviance is $D_{\mathrm{min}} + \sigma_{\mathrm{CV}}$, that is the largest $\lambda$ which is still consistent with the minimum values of $D$. This gives the most parsimonious model, containing only the predictors \lmstar, $\epsilon_{0,1}$ and \logsfr.}
    \label{fig:lasso_CV}
\end{figure}

To find the most appropriate value of $\lambda$, we split the SAMI data into 5 parts and re-fit the LASSO model on each. Figure \ref{fig:lasso_CV} shows the results of $\lambda$ against the Binomial Deviance, our chosen goodness-of-fit statistic.

The minimum deviance occurs at $\lambda_{\mathrm{min}}=0.00156\pm0.0145$, where the uncertainty is derived using our cross-validation routine. We choose to select a value of $\lambda = \lambda_{\mathrm{min}} + \sigma_{\mathrm{CV}} = 0.0161$ for our "best" LASSO model, i.e. we use the largest value of penalisation which still gives a deviance within one standard deviation of the minimum.

At this value of $\lambda$, there are three non-zero predictors in the model: \lmstar, $\epsilon_{0,1}$ and \logsfr. This is in good agreement with Section \ref{sec:bestfitpredictors}, where we found that the various information criteria preferred models with between three and four non-zero variables (these three with the addition of \logre). Overall, we find that using regularised logistic regression gives similar conclusions to the best-subset approach taken in Section \ref{sec:bestfitpredictors}.

\bsp	
\end{document}